\documentclass[journal=jpcbfk,manuscript=article]{achemso}

\usepackage{natmove}

\author{Caroline Desgranges}
\author{Jerome Delhommelle}
\email{jerome.delhommelle@und.edu}
\affiliation[University of North Dakota]
{Department of Chemistry, University of North Dakota, Grand Forks, ND 58202, USA\\
(Phone: 701-777-2495)}

\title{Effect of the Composition on the Free Energy of Crystal Nucleation for CuPd Nanoalloys.}

\begin{document}

\begin{abstract}
Using molecular simulation, we study the nucleation process from supercooled liquid alloys of Cu and Pd. The simulations reveal a complex interplay between the size of the crystal nucleus and its composition that greatly impacts the crystallization process on the nanoscale. In particular, we find that the free energy of nucleation strongly depends on the composition of the alloy, with a free energy barrier for the equimolar alloy that is more than two times larger than for the pure metals. We attribute this increase in free energy to the variations in composition occurring both at the surface and in the core of the nucleus. The local changes in composition are then analyzed by comparing the surface energies of the two metals and by taking into account the competition between crystallization and demixing that takes place at the interface between the nucleus and the surrounding liquid.
\end{abstract}

\section{Introduction}

Crystal nucleation is an incredibly complex~\cite{sosso2016crystal,bernstein2002polymorphism,tenWolde,Auer,wallace2013microscopic,kawasaki2010formation,desgranges2011role,desgranges2014unraveling,tane2016two,kovcar2016design,taudt2015simulation,de2016potential,haji2015direct,jungblut2016pathways,metya2016ice,lupi2016pre,espinosa2016seeding,mithen2015nucleation,price2015molecular,berryman2016early,statt2015finite,giberti2015metadynamics}, yet ubiquitous, phenomenon, common to many applications in materials sciences. Its complexity stems from the constant competition between thermodynamics and kinetics~\cite{ostwald1897studien} that arises during the formation of a nanoparticle and dictates the size, the crystal structure, the shape, and, ultimately, the properties of the nanomaterial. As a result, the mechanisms, at the atomic scale, accounting for the formation of metallic and bimetallic nanoparticles have remained elusive. This, in turn, limits our ability to control the properties of nanoparticles~\cite{zheng2009observation}, which are increasingly used in applications as e.g. drug carriers~\cite{agasti2009photoregulated}, DNA/protein markers~\cite{giljohann2010gold} and for high-density magnetic recording~\cite{sun2000monodisperse}. In recent years, bimetallic nanoalloys~\cite{ferrando2008nanoalloys} and nanoparticles~\cite{rossi2005global} have become the focus of intense research, as synergistic effects between the two metals often lead to enhanced catalytic properties for bimetallic nanoparticles. These systems present the additional challenge that the composition of the crystal nucleus becomes also a key parameter, in addition to e.g. the size and structure of the crystal nucleus, for the rationalization of the nucleation process.

 The aim of this work is to use molecular simulation to shed light on the mechanisms underlying the formation of nanosized crystal nuclei from supercooled liquids of CuPd alloys. We use a many-body force field, known as the quantum corrected Sutton Chen - Embedded Atoms Model (qSC-EAM)~\cite{Luo}, in our simulations of CuPd alloys. This force field has been previously shown to model accurately the properties of crystal phases of the Cu and Pd metals and of several CuPd alloys, as well as the melting temperatures for these systems~\cite{Luo,kart2008molecular}. To analyze the impact of the composition of the alloy on the nucleation process, we determine the free energy surface associated with the formation of the crystal nucleus over the entire range of compositions for the alloy. For this purpose, for each composition, we perform a series of umbrella sampling simulations~\cite{tenWolde,Auer,desgranges2014unraveling} to promote the onset of crystalline order and the formation of a crystal nucleus. We then carry out a detailed analysis of the structure, the size as well as the composition of the crystal nucleus for each composition, and focus on elucidating the processes that occur at the surface of the nucleus, i.e at the interface of the crystal nucleus with the surrounding liquid, and in the core of the nucleus. This leads us to rationalize the nucleation process in these binary systems in terms of the competition that arise between ordering and demixing as observed during the crystallization of polymer mixtures~\cite{tanaka1989local,jin2013kinetics} and, more recently, in metal alloys and nanoalloys~\cite{desgranges2014unraveling,nandi2016composition}.

 The paper is organized as follows. In the next section, we present the qSC-EAM force field used in the simulations to model the CuPd alloys. We also detail the method with which we perform the simulations for the crystal nucleation process, the structural analysis carried out to determine the structure and composition of the crystal nucleus, both at the surface and in the core of the nucleus. We then discuss the results for the free energy of nucleation and its dependence upon the composition of the CuPd alloy. We also present the results obtained for the evolution in size, structure and composition of the crystal nucleus as nucleation proceeds, and analyze the impact of the properties for the two components of the alloy, i.e. their relative sizes as well as the relative values for their cohesive and surface energies, on the characteristics of the nucleus. We finally draw the main conclusions from this work in the last section.

\section{Simulation methods}

\subsection{Force field for CuPd alloys}
Metal alloys are complex to model because of the many-body interactions that take place in such systems. A strategy that is often used is to employ a force field based on an embedded atoms approach (EAM)~\cite{finnis1984simple,sutton1990long,mei1991analytic,daw1983semiempirical}. This allows to simulate large systems of several thousands of atoms over the long simulation times necessary to give rise to the formation of a crystal nucleus. EAM force fields include density dependent terms and were initially parametrize to model the properties of the solid phases of metals. A very successful approach known as the quantum-corrected Sutton Chen embedded atoms models (qSC-EAM)~ \cite{sutton1990long,Luo} was shown to accurately model the properties of many transition metals. In subsequent work, the qSC-EAM potential was also used in simulations of the solid-liquid transition and of the liquid phase. The results from these simulations established that the qSC-EAM potential performed very well in predicting the melting temperatures~\cite{Luo,Caginpdni,kart2008molecular} of single component systems and binary allows as well as a large number of liquid properties. In this work, we use the qSC-EAM potential to study the crystallization behavior in CuPd nanoalloys for different compositions of the mixture. Interactions between atoms according to the qSC-EAM force field are calculated as the sum of a two-body contribution and of a many-body term 

\begin{equation}
U={\sum_{i=1}^N \left(  \sum_{j \ne i} {1 \over 2} \varepsilon_{ij} V(r_{ij}) - c_i \varepsilon_{ii}(\rho_i)^{1/2} \right)}
\end{equation}
where $\rho_i$ is the local density associated with an atom $i$ and $r_{ij}$ the distance between two atoms $i$ and $j$.
The pair potential is given by
\begin{equation}
V(r_{ij}) = {\left( {a_{ij} \over r_{ij}} \right)^{n_{ij}}}
\end{equation}
and the many-body part by
\begin{equation}
\rho_i=\sum_{j \ne i}{\left( {a_{ij} \over r_{ij}} \right)^{m_{ij}}}
\end{equation}

To determine the parameters for the interactions between unlike atoms,  we use a set of combining rules which has been shown to perform very well for metal nanoalloys~\cite{Caginpdni,kart2008molecular}
\begin{equation}
\varepsilon_{ij}=\sqrt{\varepsilon_i \varepsilon_j}
\label{LB1}
\end{equation}
\begin{equation}
m_{ij}={{m_i + m_j} \over 2}
\label{LB2}
\end{equation}
\begin{equation}
n_{ij}={{n_i + n_j} \over 2}
\label{LB3}
\end{equation}
\begin{equation}
a_{ij}={{a_i + a_j} \over 2}
\label{LB4}
\end{equation}

The parameters used in this work are given in Table~\ref{Tab1}, with the cutoff distance set to twice the parameter $a$. The qSC-EAM potential, together with the combining rules, accurately models the melting temperature ($T_m$) of CuPd alloys across the entire composition range (see Table~\ref{Tab2}).  

\begin{table}[h]
\small
  \caption{\ Parameters for the qSC-EAM potential.}
  \label{Tab1}
  \begin{tabular*}{0.5\textwidth}{@{\extracolsep{\fill}}llllll}
\hline
$$ & $a $~(\AA) &  $\varepsilon (10^{-2}eV)$ & $c$ & $m$ & $n$ \\
\hline
$Cu$ & $3.6030$ &  $0.57921$ & $84.843$ & $5$ & $10$ \\
$Pd$ & $3.8813$ &  $0.32864$ & $148.205$ & $6$ & $12$ \\
\hline
  \end{tabular*}
\end{table}

\subsection{Simulating crystal nucleation in CuPd nanoalloys}

We simulate the crystallization process from supercooled liquid phases of the CuPd nanoalloy. Crystal nucleation is an activated process, as the system has to overcome a large free energy barrier of nucleation to form a crystal nucleus of a critical size. According to classical nucleation theory~\cite{mcgraw1997interfacial}, the critical size corresponds to the size necessary for two contributions, of opposite signs, to the free energy of the system to cancel out. The first, energetically favorable, contribution is the 'conversion' of the metastable phase (here, the supercooled liquid) into the stable phase (the crystal phase), while the second, unfavorable, term is associated with the cost of creating (and increasing the area of) the interface between the crystal nucleus and the surrounding liquid. As shown by Turnbull~\cite{Turnbull} , metals often exhibit a large free energy barrier of nucleation. As a result, metals and alloys remain liquid, even if they are cooled far below their melting point without crystallizing. Previous work has connected this behavior with the icosahedral order that can often be found in supercooled liquid metals and that prevents crystallization~\cite{Franck,Kelton}. The large free energy barriers of nucleation for metals and alloys implies that, even for large supercoolings, the spontaneous formation of a critical nucleus cannot take place during the time scales spanned by conventional Monte Carlo (MC) or a Molecular Dynamics (MD) simulations. These methods rely on a Boltzmann sampling scheme, and, as a result, configurations at the top of the free energy barrier, and thus, of high free energy are not sampled. Methods have been developed in recent years to study the nucleation process~\cite{Torrie,tenWolde,Moroni,Allen1,Beckham2}. The non-Boltzmann sampling technique we use in this work is known as the umbrella sampling method~\cite{Torrie}. This method consists of adding a bias potential energy to the total energy of the system, in order to promote the formation of the crystal nucleus. The bias potential energy is often chosen to be a harmonic function of a reaction coordinate, which measures the progress of the system towards the formation of a crystal nucleus. In this work, we choose as the reaction coordinate the order parameter~\cite{Steinhardt} $Q_6$ as the reaction coordinate. Previous work on $Q_6$ has shown that it is extremely well-suited to study the onset of crystalline order as discussed in experiments on colloidal suspensions~\cite{Gasser} and in simulations of the crystallization process for many systems~\cite{tenWolde,JACS1,Auer,JACS2}, including metals~\cite{Nam,JPCCAu,JACS3,JCPAl,hou2015formation,JohnFe,li2014nucleation}. $Q_6$ has also been successfully applied to the study of the crystallization in binary mixtures of simple fluids~\cite{Monson,jungblut2011crystallization}, as well as of metal alloys~\cite{PdNi,chen2016non,celik2015investigation,an2016two}. We carry out simulations of the nucleation process on systems of a total of $4,000$ atoms, for pure Cu and pure Pd, and for three different mole fractions in Pd ($x_{Pd}=0.25$, $x_{Pd}=0.5$ and $x_{Pd}=0.75$). We present in Table~\ref{Tab2} the conditions of crystallization for each of the systems studied in this work at $P=1$~bar.

\begin{table}[h]
\small
\caption{\ Melting temperatures ($T_m$) and crystallization conditions ($T_{cryst}$ corresponding to a supercooling of $35$~\%) for the different CuPd systems.}
\label{Tab2}
\begin{tabular*}{0.8\textwidth}{@{\extracolsep{\fill}}llllll}
\hline
$ x~(Pd) $ & $0$ &  $0.25$ & $0.5$ & $0.75$ & $1$ \\
\hline
$T_{m}~(exp)$~\cite{hultgren1973selected} & $1356$ &  $1405$ & $$ & $1650$ & $1825$ \\
$T_{m}~(sim)$~\cite{kart2008molecular} & $1370$ &  $1450$ & $1575$ & $1700$ & $1820$ \\
$T_{cryst}$ & $890.5$ &  $942.5$ & $1024$ & $1105$ & $1183$ \\
\hline
\end{tabular*}
\end{table}

Throughout the nucleation process, we carry out a detailed structural analysis of the system to identify which atoms belong to the incipient crystal nucleus and what type of crystal structure forms as nucleation proceeds. For this purpose, we first determine the type of environment (liquid-like or crystal-like) around each atom by keeping track of the correlation between the bond vectors associated with each atom (in this work, we use a cutoff radius of $3.3$~\AA~to determine that two atoms are nearest neighbors). Once an atom has been identified as having a crystal-like environment, we assign a specific crystal structure on the basis of another local parameters~\cite{Steinhardt}, $w_6(i)$ for each atom $i$. As in previous work, a crystal-like atom $i$ is labelled as body-centered cubic ($BCC$) if $w_6(i)$ is greater than $0$ and is labelled as closed-packed structure ($CP$) otherwise. We also looked for other types of crystal structures including the ordered phases $L_{12}$ and $B_2$ which have been reported at low temperature for the $Cu_3Pd$ (or $CuPd_3$) and for the CuPd crystals, respectively. However, our analysis only revealed disordered crystal phases, which is why we only discuss the crystal structure of the nuclei in terms of the $CP$ and $BCC$ structures. Finally, we also analyze the evolution of the nucleus in terms of its core and surface atoms. Core atoms are identified as atoms for which all of the nearest neighbors have a crystal-like environment. The remaining atoms of the nucleus are then defined as surface atoms.

\section{Results and discussion}

\begin{figure}
\begin{center}
\includegraphics*[width=8cm]{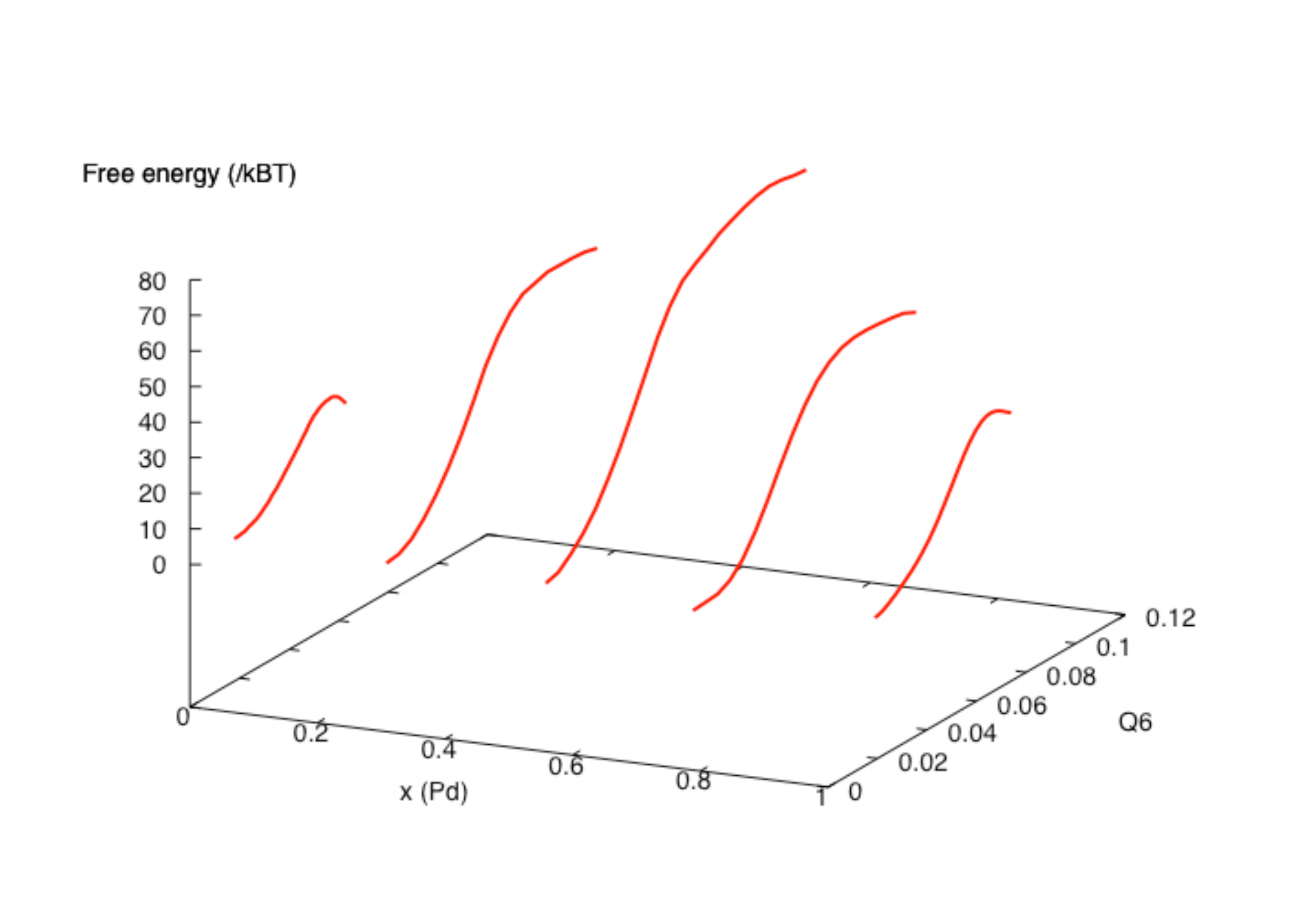}
\end{center}
\caption{Free energy of nucleation (/$k_BT$) of CuPd nanoalloys against the mole fraction in Pd, $x_{Pd}$, and the order parameter $Q_6$. The plot shows that the free energy increases as we move towards a mole fraction of $0.5$.}
\label{Fig1}
\end{figure}

We start by discussing the results obtained for the free energy of nucleation of the CuPd nanoalloys. Fig.~\ref{Fig1} shows a plot of the free energy barrier of nucleation as a function of the reaction coordinate $Q_6$ for $x_{Pd}=0$ (pure Cu), $x_{Pd}=0.25$, $x_{Pd}=0.5$, $x_{Pd}=0.75$ and $x_{Pd}=1$ (pure Pd). We observe the following main trends. First, for any value of $x_{Pd}$, the free energy of the system always starts to increase with $Q_6$ before a maximum is reached. This increase corresponds to the formation of a small crystal nucleus, which is an unfavorable process due to the predominant cost of creating the solid-liquid interface. Then, the free energy profile of nucleation reaches a maximum associated with the formation of the critical nucleus. The value of the order parameter $Q_6^c$ for which the critical nucleus has formed depends on the Pd mole fraction with e.g. $Q_6^c=0.105$ for $x_{Pd}=0.5$ while it is of about $Q_6^c=0.08$ for $x_{Pd}=0.25$ or $x_{Pd}=0.75$ and of about $Q_6^c=0.06$ for the pure metals. The second feature of the free energy plot is the increase of free energy as the alloy starts to approach an equimolar composition. This can best be seen in the case of $x_{Pd}=0.5$, for which the top of the free energy barrier is reached for a free energy of $75\pm 5~k_BT$. This is about $40$\% more than the free energy of nucleation for $x_{Pd}=0.25$ and $x_{Pd}=0.75$ and more than twice the free energy of nucleation for the pure metals. This result is consistent with our findings for the $Q_6^c$ values which are markedly greater for the alloys than for the pure metals. A greater value for a $Q_6^c$ means that the system has to reach a greater level of organization, by forming a larger crystal nucleus and, as a result, the system will have to overcome a larger free energy barrier of nucleation to achieve this. This phenomenon is even more pronounced as we reach the equimolar composition for the alloy, resulting in the highest free energy barrier obtained in the middle of Fig.~\ref{Fig1}. We add that this result is consistent with previous work on the AgPd mixture, which also revealed an increase in the free energy barriers of nucleation for AgPd alloys~\cite{desgranges2014unraveling}.

Several reasons can potentially account for this steep increase of the height of the free energy barrier for the alloy. A first possible explanation lies in the difference between the sizes for the two atoms, since we have $a=3.603$~\AA~for Cu and $a=3.8813$~\AA~for Pd. The difference between the sizes of the two types of atom composing the alloy may result in a mismatch during the nucleation process and hinder the formation of the thermodynamically preferred $CP$ structure, potentially leading to higher free energy barrier of nucleation for the alloy. The second factor leading to the greater barrier observed for the alloys, may result from the energetics for the formation of the crystal phase. To examine this, we consider the following two quantities, the cohesive energy and the surface energies. First, comparing the cohesive energies for Cu and Pd, we find that the qSC-EAM potential predicts a larger cohesive energy for Pd ($3.89~eV$) than for Cu ($3.49~eV$). This means that the 'conversion' of liquid-like Pd atoms into crystal-like Pd atoms will result in a greater energetic gain. This can lead to the preferred incorporation of Pd atoms into the crystal nucleus during the nucleation process. Such a selection of the composition of the nucleus during its formation will also lead to a larger free energy barrier of nucleation. Finally, considering now the relative values for the surface energies, we find that for Cu, depending on the crystallographic plane considered~\cite{Luo}, the qSC-EAM potential predicts surface energies which are between $1495$ and $1657$~$mJ/m^2$, while for Pd exhibits lower surface energies (between $1229$ and $1416$~$mJ/m^2$). This implies that the crystal nucleus will have a lower surface energy if more Pd atoms are located at the surface, a factor that will impact the interplay between composition and size of the crystal nucleus.

\begin{figure}
\begin{center}
\includegraphics*[width=8cm]{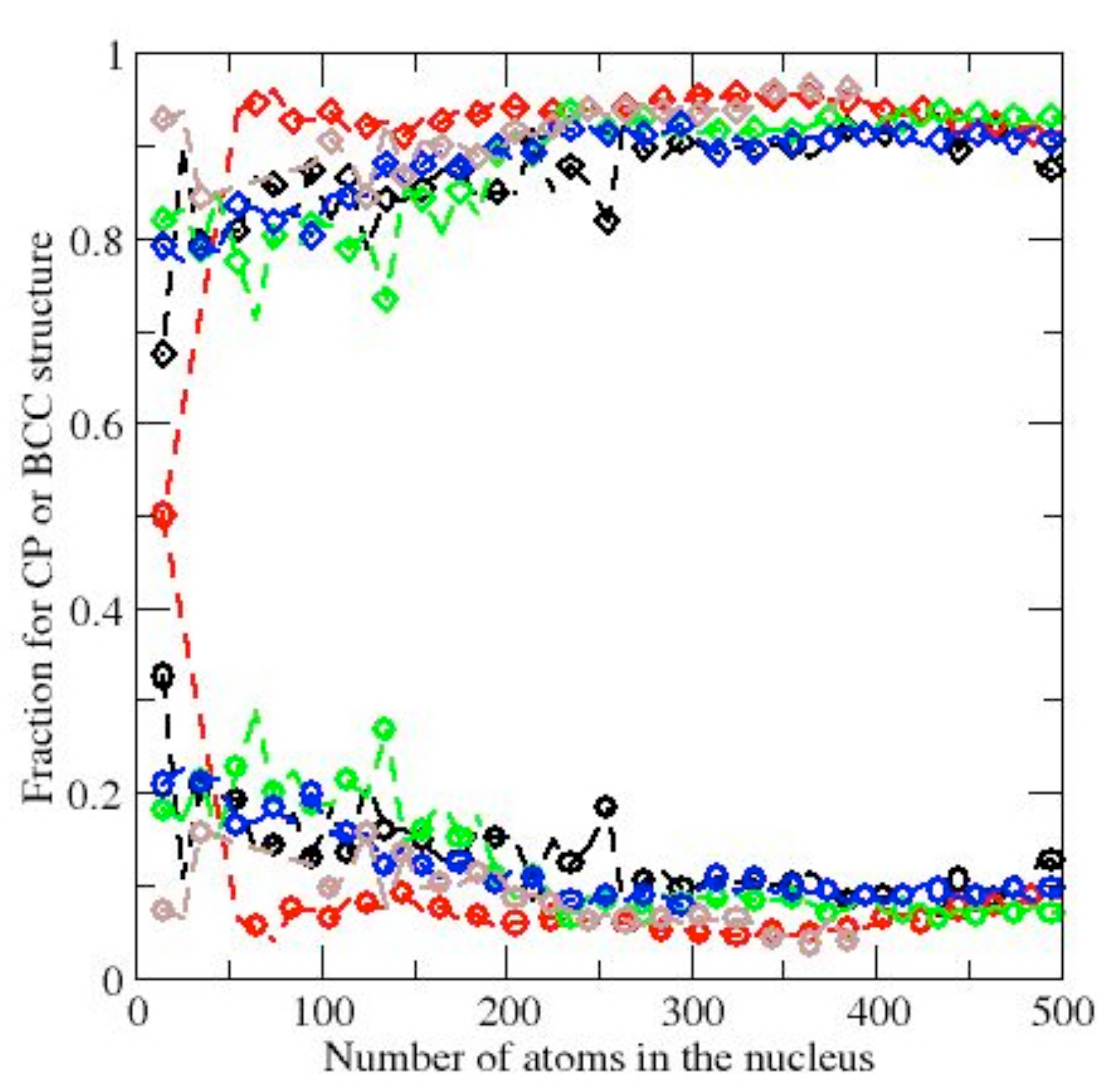}
\end{center}
\caption{Fraction of the closed-packed (CP) structures (diamonds) and the body-centered cubic (BCC) structure (circles) in the nucleus as a function of its overall size. We show the results for the nanoalloy with $x_{Pd}=0$ in black, for $x_{Pd}=0.25$ in blue, for $x_{Pd}=0.50$ in green, for $x_{Pd}=0.75$ in red and for $x_{Pd}=1$ in brown.}
\label{Fig2}
\end{figure}

\begin{figure}
\begin{center}
\includegraphics*[width=8cm]{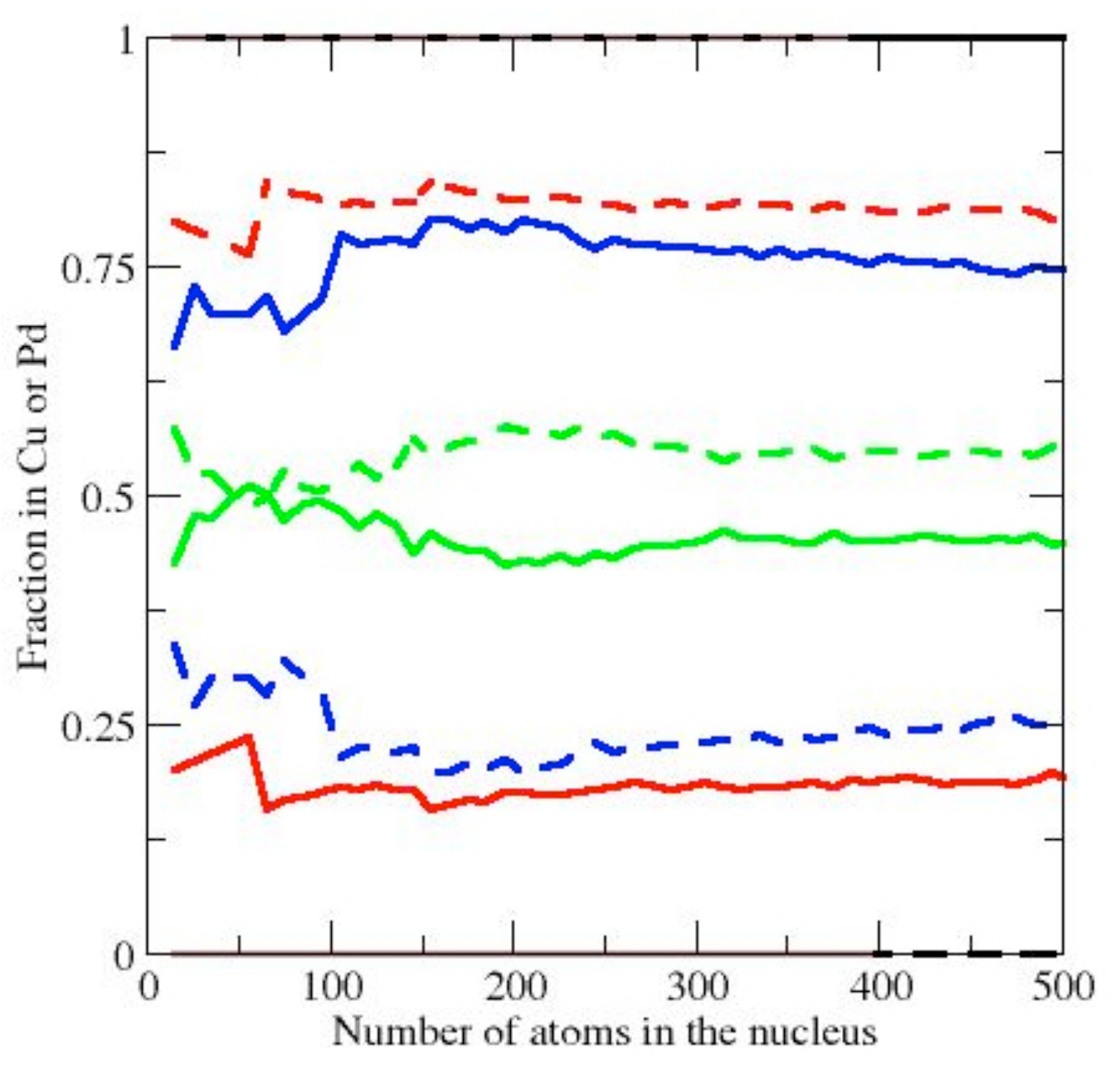}
\end{center}
\caption{Composition of the crystal nucleus as a function of its size (Cu: solid line and Pd: dashed line). The colors for each composition are assigned as in Fig.~\ref{Fig2}.}
\label{Fig3}
\end{figure}

To identify the reasons underlying the steep increase in the free energy barrier for the alloy, we carry out a detailed analysis of the structure and composition of the nucleus throughout the nucleation process. We show in Fig.~\ref{Fig2} the evolution of the fraction for each crystal structure, and in Fig.~\ref{Fig3}, the evolution of the composition of the nucleus as nucleation advances. We add that these two plots display the overall behavior observed for the entire (surface+core) nucleus. Nucleation starts with the formation of a crystal nucleus which is predominantly $CP$ but also contains a significant amount of atoms with a $BCC$ signature (of the order of $20$~\% for nuclei containing less than $50$ atoms). The $BCC$ fraction, however, decreases very quickly during nucleation and stabilizes around $10$~\% for all systems, when the nucleus reaches a size in excess of $200$ atoms. These results indicate that regardless of the mole fraction in Pd, the structure of the nucleus is predominantly $CP$, which tends to show that the structural mismatch due to the size difference between Cu and Pd atoms  does not impact the overall structure of the nucleus and, as such, that the main cause for the increase in the free energy of nucleation for the alloys is not due to a structural difference with the crystal nuclei obtained for pure metals. 

We now turn to the variations in the composition of the crystal nucleus during the nucleation process. Fig.~\ref{Fig3} shows that, during the early stages of the nucleation process, the fraction in Pd in the nucleus is always greater than in the alloy. For instance, for the alloy with $x_{Pd}=0.25$, we find a fraction in Pd that remains above $30$~\% for all sizes of the nuclei below $100$ atoms. Similarly, for the alloy $x_{Pd}=0.5$, nuclei of less than $50$ atoms are richer in Pd. Even when the value for $x_{Pd}$ is very large ($x_{Pd}=0.75$), we obtain a nucleus that is predominantly composed of Pd atoms for small sizes ($N < 50$). At this stage of the nucleation process, the nucleus is mainly composed of atoms located at the surface and therefore, since Pd atoms have a lower surface energy, the nucleus has a high Pd fraction. Then, we observe for all alloys that the composition changes very quickly once its size exceeds $50-100$ atoms with a drop in the fraction in Pd in the nucleus. This is likely due to the nucleus growing and incorporating the neighboring Cu atoms from the liquid in the immediate vicinity of the crystal nucleus. This shows that a competition between ordering and demixing occurs during the nucleation process. Then, we observe an increase in the mole fraction in Pd in the nucleus (see e.g. the top of Fig.~\ref{Fig3} for $x_{Pd}=0.75$). At the end of the nucleation process, we find that the mole fraction in Pd for the nucleus is either very close to that of the bulk ($x_{Pd}=0.25$) or higher, with a Pd fraction in the nucleus of $0.55$ for the system $x_{Pd}=0.5$ and of $0.8$ in the nucleus for $x_{Pd}=0.75$. This result can be attributed to the larger cohesive energy of Pd atoms which accounts for the larger fraction in Pd for larger nuclei. We add, however that the excess of Pd in the crystal nucleus does not significantly impact the Pd mole fraction in the remaining liquid. This is due to the small size of the critical nucleus when compared to the overall system size. For instance, for $x_{Pd}=0.5$, the Pd fraction in the remaining liquid is of $49.6$~\% while it is of $74.3$~\% when $x_{Pd}=0.75$. To ascertain these conclusions, we now move on to the analysis of the core and of the surface of the crystal nuclei. 

First, we examine the structure of the atoms at the surface of the nuclei for the different composition of the alloys. We observe some subtle deviations from the behavior discussed in the case of the entire nuclei. As for Fig.~\ref{Fig2}, we find that the atoms at the surface are more $BCC$-like at the beginning of the nucleation process than at the end. However, the evolution of the structure of the surface towards the $CP$ structure is much slower and much more gradual. This can be attributed to the greater amount of disorder and the greater presence of defects at the interface between the nucleus and the surrounding liquid. This behavior is qualitatively the same for all compositions of the alloys. 

\begin{figure}
\begin{center}
\includegraphics*[width=8cm]{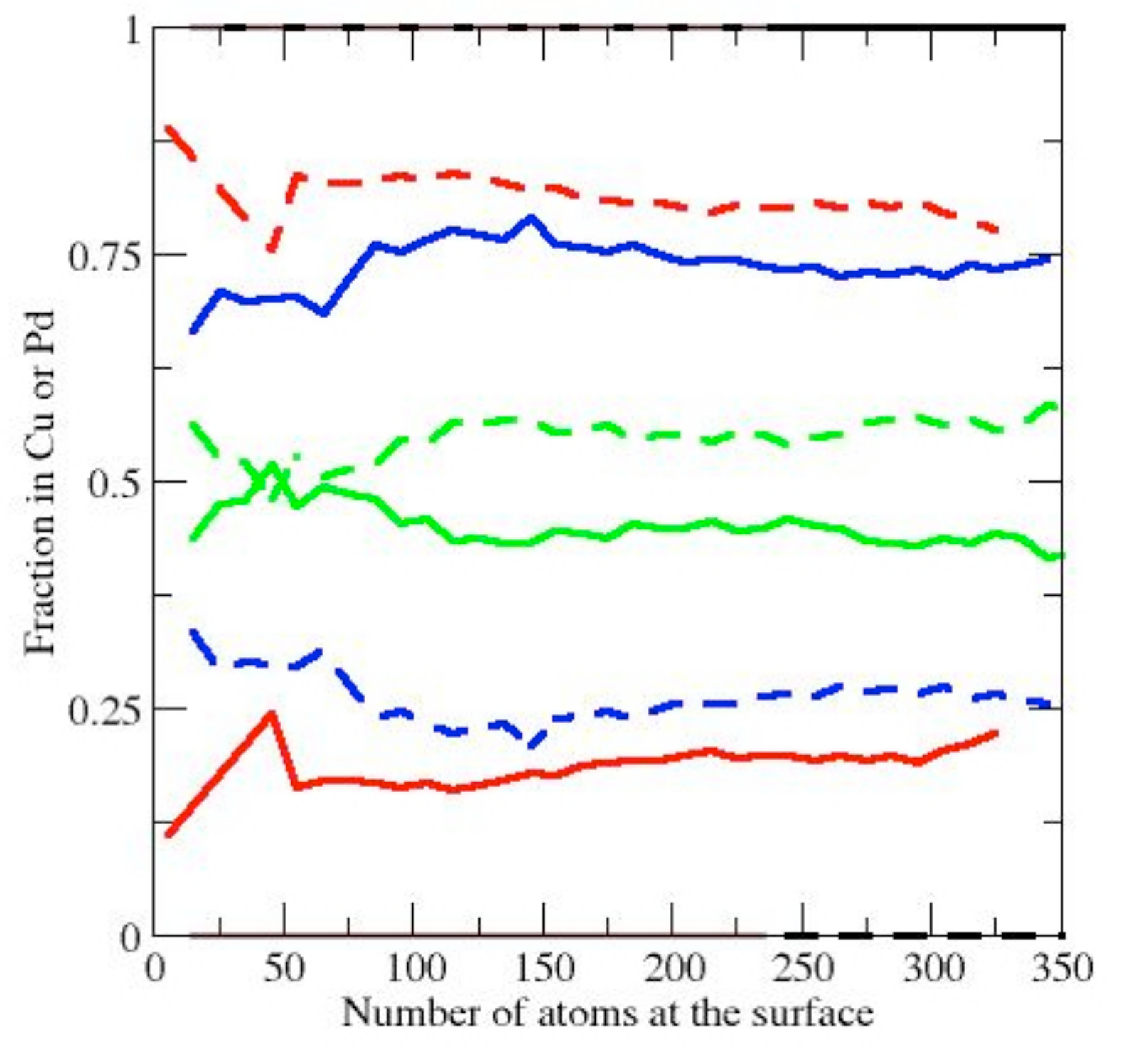}
\end{center}
\caption{Mole fractions of Cu (solid line) and of Pd (dashed line) as a function of number of atoms located at the surface of the crystal nucleus.}
\label{Fig4}
\end{figure}

The composition at the surface of the nucleus closely mirrors the overall composition of the nucleus during the early stages of the nucleation process. As a result, the trend observed on the left of Fig.~\ref{Fig4} is in line with that observed for small nuclei in Fig.~\ref{Fig3}, with nuclei forming first with an excess of Pd (when compared to the composition of the bulk). This is then compensated by the inclusion of the Cu atoms from the surrounding liquid. We then see that the composition of the surface generally tends to be richer in Pd than its counterpart for the entire nucleus, which confirms the predominant role played by the lower surface energy of Pd in the nucleation process.

The structure associated with the atoms located at the core of the crystal nucleus shows some notable differences with that observed for the surface. In particular, even early in the nucleation process, the structure of the core is much more of the $CP$ type. Very rapidly, the fraction of the $CP$ structures exceeds $90$~\% (for a number of atoms greater than $100$ in the core) and becomes greater than $95$~\% for all compositions as soon as the size of the core is beyond $150$ atoms. This means that the surface atoms that were present in small nuclei (which contain a significant fraction of the $BCC$ structure) quickly convert into the $CP$ structures as they become part of the core of the nucleus. As for Fig.~\ref{Fig2}, we observe a similar behavior for all alloy compositions which point to the fact that the mismatch in size between the two atoms does not impact strongly the structure of the core of the nuclei.

\begin{figure}
\begin{center}
\includegraphics*[width=8cm]{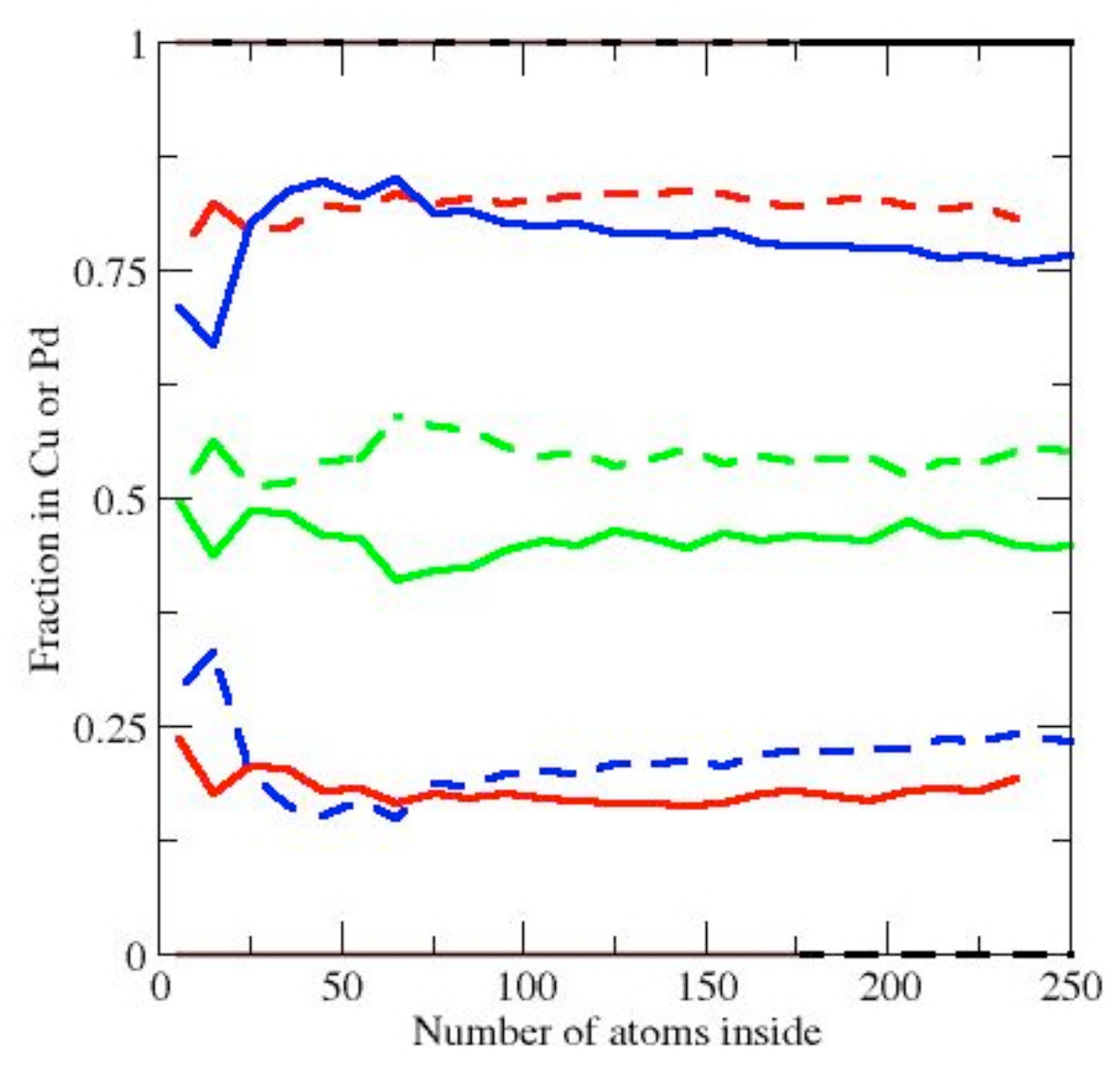}
\end{center}
\caption{Mole fractions of Cu (solid line) and of Pd (dashed line) against the number of atoms located inside the crystal nucleus.}
\label{Fig5}
\end{figure}

Fig.~\ref{Fig5} shows the variation of the mole fractions in the core as a function of the the number of atoms inside the nucleus. Since the surface of the nuclei was initially rich in Pd, we observe a core that starts predominantly with Pd atoms. This conclusion is found to hold for all alloy compositions. Then, we observe behaviors that depend on the composition of the alloy. For $x_{Pd}=0.25$, we find that the drop in the mole fraction in Pd that results from the incorporation of the Cu atoms from the liquid in the immediate vicinity, leads to a core that shows a lower concentration in Pd than for the bulk. This shows the great significance of the processes occurring at the surface of the nucleus. The system 'chooses' to keep a high fraction of Pd atoms, of lower surface energy than Cu atoms, at the surface. This process is done at the expense of the energy of the core of the nucleus, as there are fewer Pd atoms of greater cohesive energy than Cu atoms in the core. For $x_{Pd}=0.5$ and $x_{Pd}=0.75$, we observe that the drop in the Pd mole fraction observed for core sizes close to $50$ atoms does not durably impact the composition of the inside of the nucleus since its composition remains rich in Pd.  

\begin{figure}
\begin{center}
\includegraphics*[width=6cm]{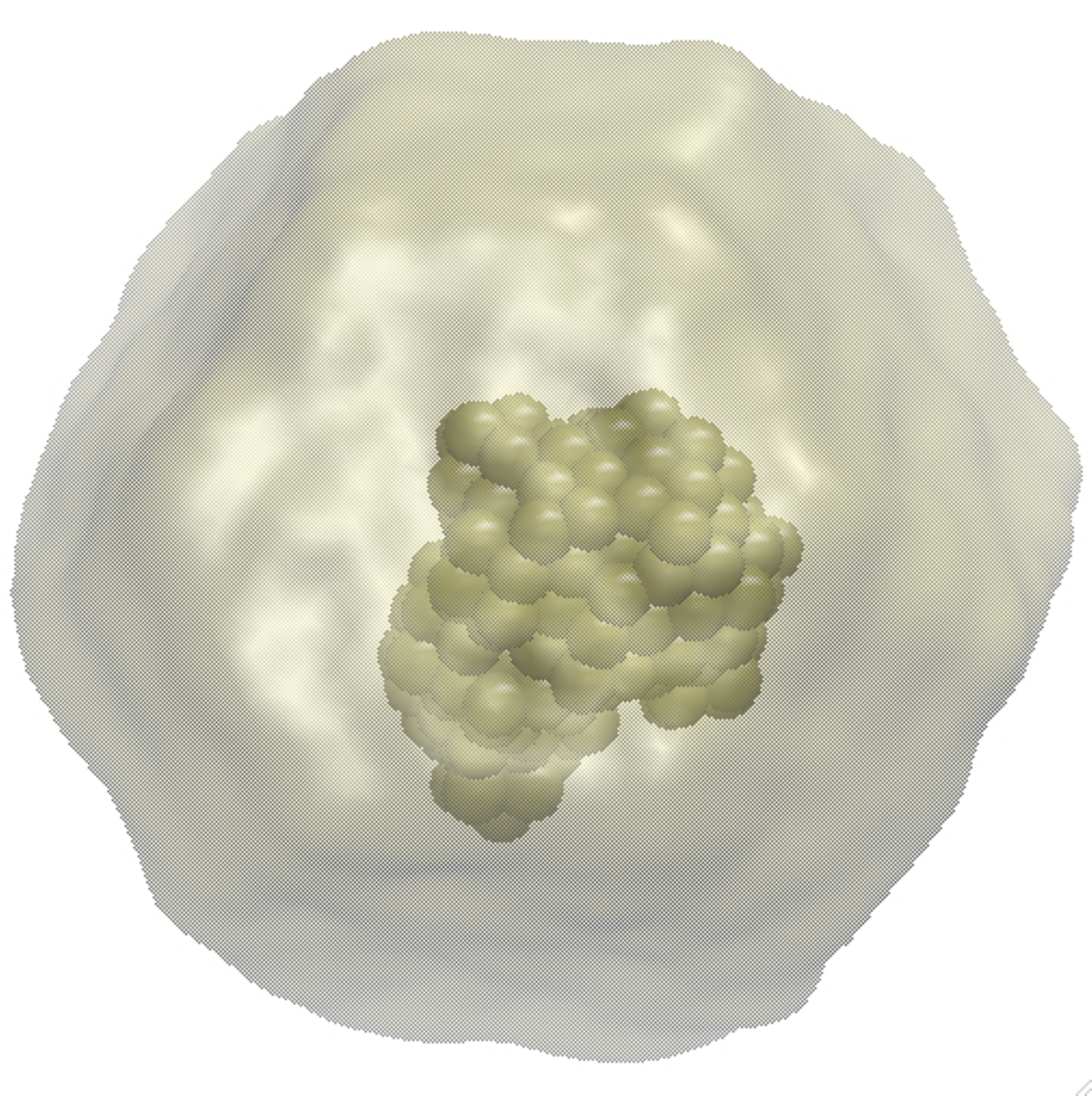}(a)
\includegraphics*[width=6cm]{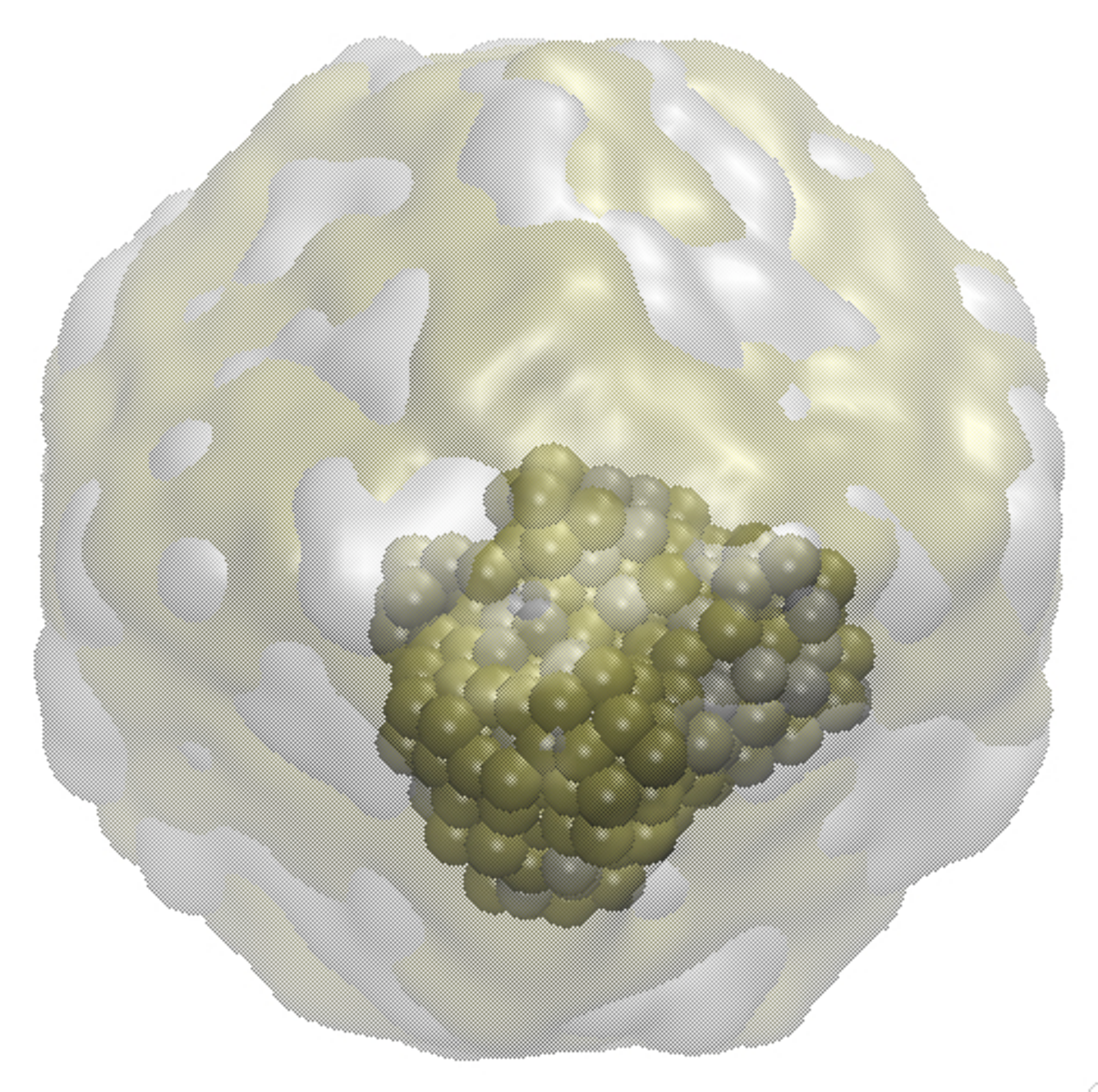}(b)
\includegraphics*[width=6cm]{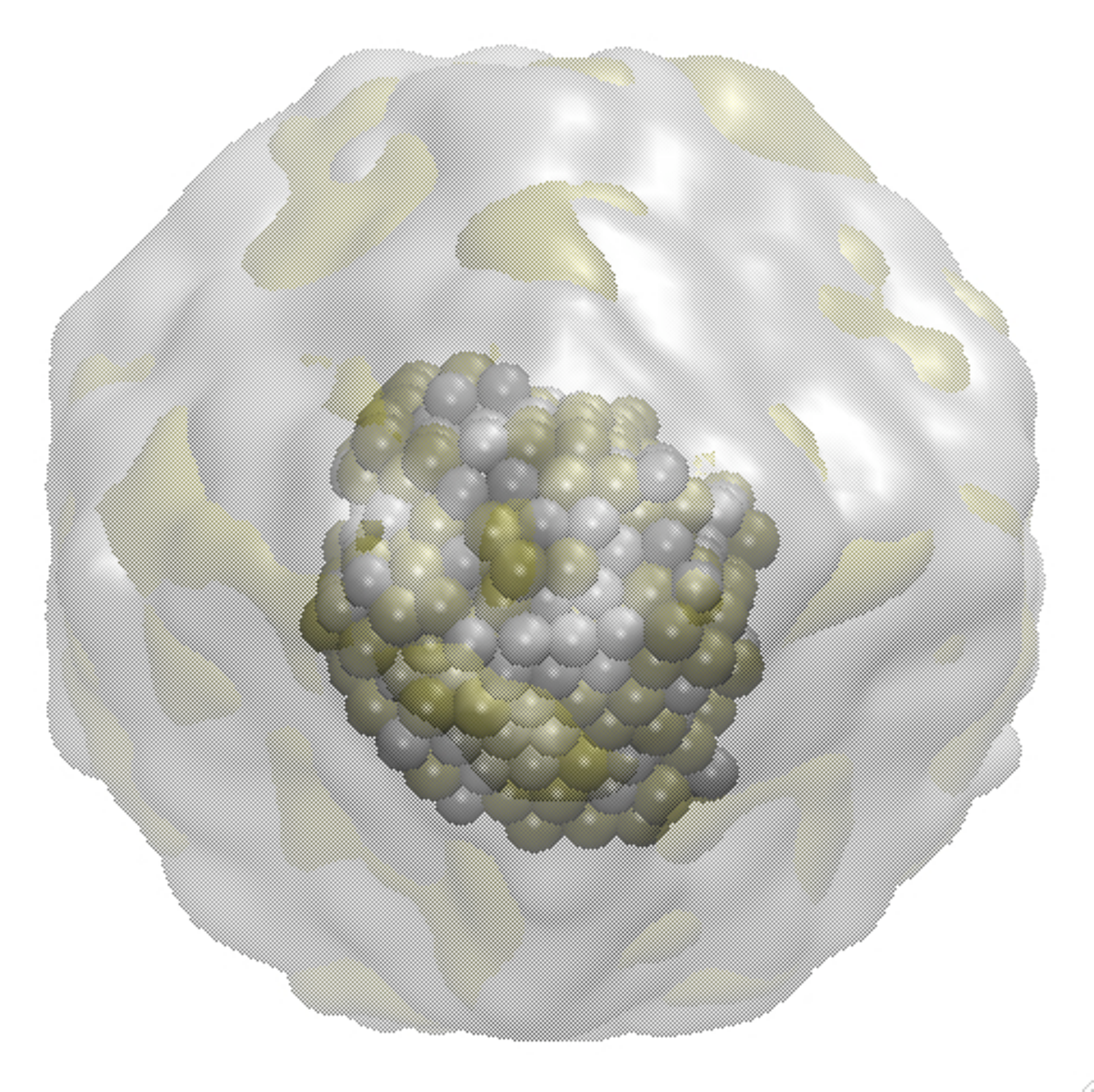}(c)
\includegraphics*[width=6cm]{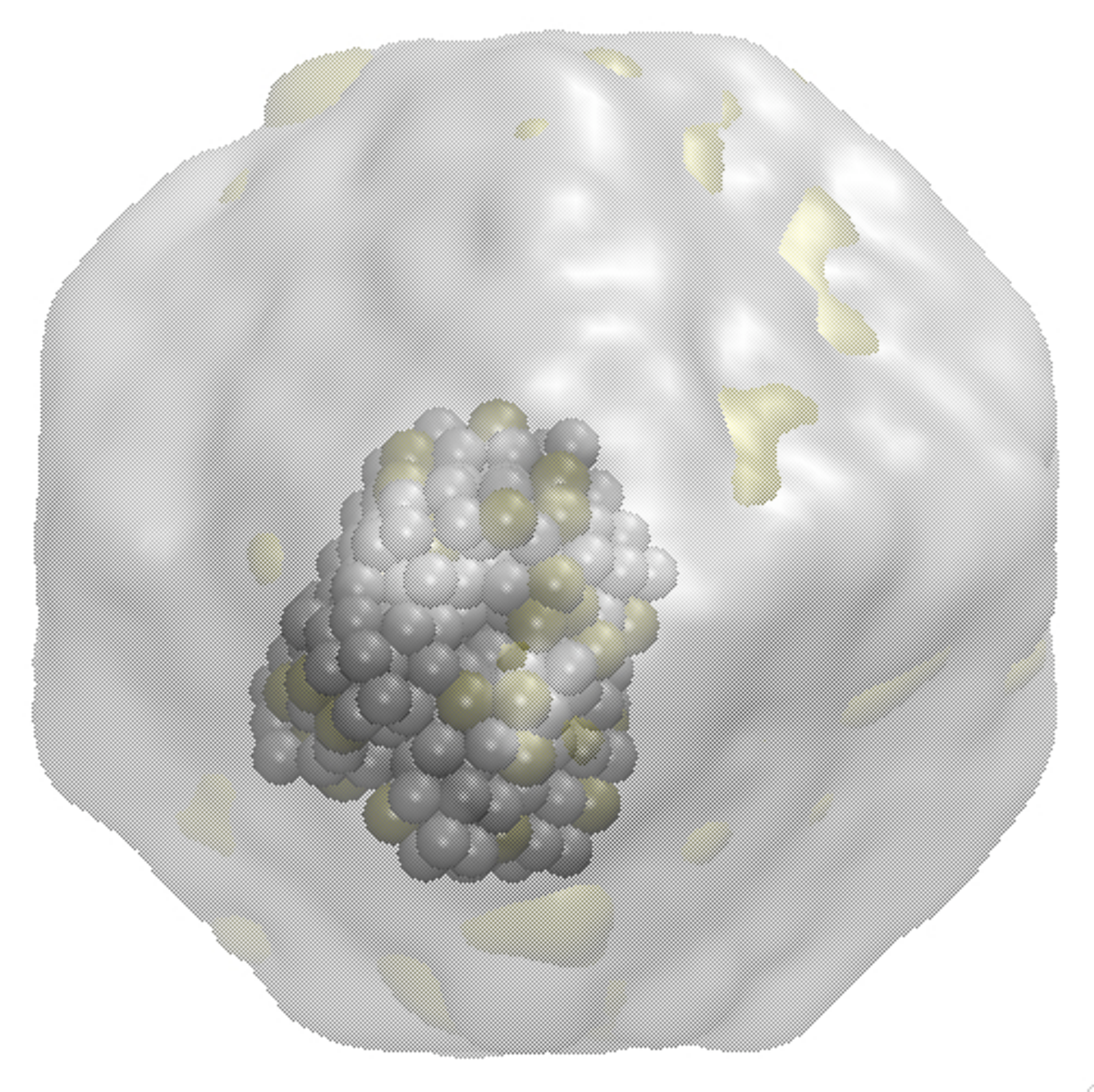}(d)
\includegraphics*[width=6cm]{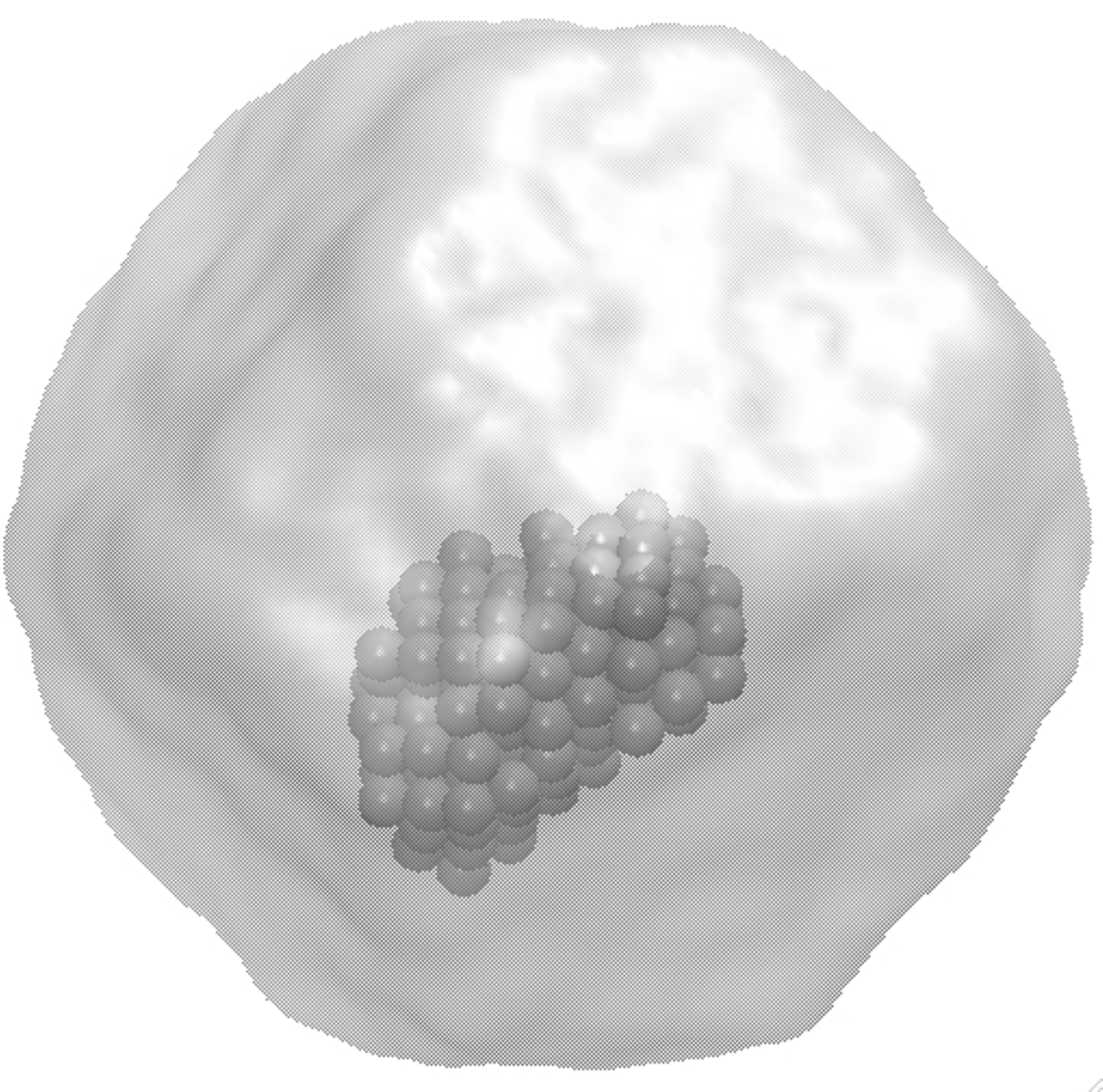}(e)
\end{center}
\caption{Snapshots of the critical nuclei obtained for each nanoalloy. (a) $x_{Pd}=0$, (b) $x_{Pd}=0.25$, (c) $x_{Pd}=0.5$, (d) $x_{Pd}=0.75$ and (e) $x_{Pd}=1$. The atoms belonging to the crystal nucleus are shown as spheres (Cu is shown in tan and Pd in silver).}
\label{Fig6}
\end{figure}

We present in Fig.~\ref{Fig6} snapshots of the critical nuclei obtained for each alloy composition. In each case, the entire simulation cell (i. e. truncated octahedron) is shown with the atoms belonging to the crystal nucleus shown as spheres. The critical nuclei are the smallest for the two pure metals, with a size of $223 \pm 39$ total atoms for Cu ($152 \pm 25$ atoms at the surface and $71 \pm 14$ atoms in the core) and $243 \pm 50$ total atoms for Pd ($154 \pm 31$ atoms at the surface and $89 \pm 20$ atoms in the core). Then for intermediate mole fractions, we obtain slightly larger critical nuclei. For $x_{Pd}=0.25$, we have a critical nucleus containing a total of $285 \pm 38$ atoms ($177 \pm 19$ atoms at the surface and $108 \pm 19$ atoms in the core). For $x_{Pd}=0.75$, we have a critical nucleus containing a total of $265 \pm 41$ atoms ($167 \pm 21$ atoms at the surface and $98 \pm 20$ atoms in the core). The largest critical nucleus is observed for the equimolar alloy with a total of $448 \pm 45$ atoms ($250 \pm 25$ atoms at the surface and $192 \pm 20$ atoms in the core). The increase in size for the critical nucleus as we approach the equimolar composition, is correlated with the increase in the free energy of nucleation. It is also consistent with the greater values for $Q_6^c$ obtained at the top of the free energy barrier for CuPd alloys when compared to the pure metals.

\section{Conclusions}
In this work, we have calculated the free energy of crystal nucleation for CuPd alloys and determined its dependence on the composition of the alloy. Our simulations reveal that the free energy of nucleation increases significantly as the mole fraction in Pd approaches $0.5$ with e. g. the alloy with $x_{Pd}$ being associated with a free energy barrier $40$~\% higher than the alloys with $x_{Pd}=0.25$ or $x_{Pd}=0.75$, and more than twice that for the pure metals. We interpret this increase in the free energy of nucleation in terms of their relative properties, i.e. the relative size of the atoms composing the alloy and their relative surface and cohesive energies. The simulations shed light on the predominant role played by the processes occurring at the surface of the developing crystal nucleus. In particular, the results show that the composition at the surface of the nucleus varies significantly during the early stages of the nucleation process, starting with nuclei rich in the element of lower surface energy (Pd). Then, as a result of the depletion in Pd that occurs for the liquid in the immediate vicinity of the crystal nucleus, the composition at the surface of the nucleus undergoes a sharp change with the rapid incorporation of Cu atoms in the nucleus. These variations in composition continue throughout the nucleation process, illustrating the competition between crystallization and demixing in the liquid alloy during the nucleation process. This has important consequences on the composition of the nanoparticle, both at the surface and in its core, and will likely impact its properties for applications e.g. in catalysis.

\acknowledgement
Partial funding for this research was provided by NSF through CAREER award DMR-1052808.\\

\bibliography{Critical}

\end{document}